\documentstyle[12pt,epsf]{article}

\begin{document}
\topmargin=-5mm
\title{The Ising Model for Changes in  Word Ordering Rule \\
in Natural Languages} 
\author{Yoshiaki Itoh and  Sumie Ueda\\
The Institute of Statistical Mathematics\\
4-6-7 Minami-Azabu, Minato-ku, Tokyo 106-8569, Japan}
\date{}
\maketitle

\abstract{
The order of `noun and adposition' is the important parameter
 of word ordering rules in the world's languages. The seven 
parameters, `adverb and verb' and others, have a strong dependence 
on the `noun and adposition'.  Japanese as well as Korean, Tamil 
and several other languages seem to have a stable structure of 
word ordering rules, as well as   Thai and other languages which  have  
the opposite 
word ordering rules to Japanese.   
It seems that each language 
 in the world
fluctuates between these  two structures like the Ising model for finite
lattice. }

\vspace{2ex}

\noindent
{\bf   Model of  changes of  word ordering rules and Tsuonoda's table}

Natural languages have  been    attractive objects of various 
interdisciplinary studies.   The problem
how the  
universals of languages evolve is  drawing the
 attension of the physicists~\cite{Gell}.    
The importance of adpositions 
(prepositions and postpositions) as a parameter is recognized 
in word order typology~\cite{Greenberg,Hawkins}.     
  In previous studies~\cite{Tsunoda1, Tsunoda2}    
using the word order data of Tsunoda (Table.1)~\cite{Tsunoda1}, in which 
Japanese is taken as the standard of comparison,    
130 languages are classified by using 19 word order parameters.  
The result is that the 130 languages are neatly divided into two groups  
with one exception or two: (a) prepositional languages and 
(b) other languages, i.~e. postpositional languages and adpositionless
languages.  Adpositionless languages behave like postpositional 
languages in terms of other word order parameters.  
The second important parameter of word order is `numeral and
noun'~\cite{Ueda}.  Applying a statistical method for categorical
data~\cite{Sakamoto}, 
the clustering of the 130 languages 
is well explained by the two word ordering parameters, 
`noun and adposition' and `numeral and noun'.
     
\begin{table}[hbtp]
\caption{Word order data by Tsunoda (1991), :  
From parameter 2  `noun and adposition' to parameter 10 `Main 
verb and auxiliary verb', and also for parameters 18 
`Conditional clause and main clause' and 19 
`Purpose clause and main clause', 
the plus sign `$+$' and minus sign `$-$' 
are used --- wherever possible --- with Japanese as the standard 
of comparison.  
(This `$+/-$' method is used for some other parameters as well when 
applicable.) 
Japanese is convenient as the standard of comparison.  
   Thus, if a given language has `noun+postposition' like Japanese, 
then it will be assigned `$+$' for parameter 2.  
If a given language has `Preposition+noun' in contrast with Japanese, 
it will be assigned `$-$' for this parameter.  
If a given language has some other 
order, then an explanation is given as much as possible.  If such 
an explanation is not feasible, it is simply presented as `Other'.  
If a given language has two orders, such as `Adjective+noun ($+$)' and 
`noun+adjective ($-$)', then the order that appears to be the more 
common is presented first.  Thus, if `Adjective+noun' is more 
common than `noun+adjective' in a given language, we have `$+/-$' 
rather than `$-/+$' for the parameter `Adjective and noun' of this 
language.
When a given language has alternative possibilities other
than the word order listed, this is shown with `etc.'.
The expression `NA (not available)' indicates that no information is available. 
}

\vspace{2ex}
{\footnotesize
\begin{tabular}{rl|l|l|l|l}
No.  &Word order parameters  &\multicolumn{1}{c|}{Japanese}
&\multicolumn{1}{c|}{English} &\multicolumn{1}{c|}{Thai}
&\multicolumn{1}{c}{Panjabi}\\[0.5ex]
\hline
1.   &S, O and V                  &SOV, etc.        &SVO        &SVO &SOV\\
2.   &Noun and adposition           &+              &$-$        &$-$ &+\\
3.   &Genitive and noun             &+              &+, $-$     &$-$ &+\\
4.   &Demonstrative and noun        &+              &+          &$-$ &+\\
5.   &Numeral and noun              &+              &+          &$-$ &+\\
6.   &Adjective and noun            &+              &+          &$-$ &+\\
7.   &Relative clause and noun      &+              &$-$        &$-$ &other; +\\
8.   &Proper noun and common        &+              &$-$, +     &$-$ &NA\\
     &noun                                                    &&&  &(not available)\\
9.   &Comparison of superiority    &+              &$-$        &$-$ &+\\
10.  &Main verb and auxiliary       &+              &$-$        &$-$, + &+ \\
     &verb     &&& \\
11.  &Adverb and verb               &before V       &various &various
&immediately \\
     &&&&       &after S\\
12.  &Adverb and adjective          &+              &+, $-$     &$-$ &+\\
13.  &Question marker   &sentence-final &none   &sentence-final,&none \\
     &                  &               &       &immediately after\\
     &                  &               &       &focus of question\\
14.  &S,V inversion in general      &none           &present  &none &none\\
     &questions &&&\\
15.  &Interrogative word   &declarative &sentence-initial&declarative
&immediately \\
     &&sentence type &&sentence type &before verb \\
16.  &S,V inversion in special      &none           &present &none &none\\
     &questions &&&\\
17.  &Negation marker              &verbal suffix  &immediately &immediately
before &immediately \\
     &&&after verb&focus of negation &before verb\\
18.  &Conditional clause and            &+      &+, $-$         &$-$, + &+\\
     &main clause&&&\\
19.  &Purpose clause and main           &+      &$-$            &$-$ &$-$\\
     &clause&&&\\
\end{tabular}
\label{order}
}
\end{table}

 Greenberg~\cite{Greenberg} utilizes 
the position of adpositions in seven of the forty five universals he 
proposes (pp.110-12). For example in languages with prepositions, 
the genitive almost 
always follows  the governing noun, while in languages with 
postpositions it almost always precedes it. 
 Tsunoda's data show, the   parameters  
1 (S, O and V), 3 (genitive and noun),  
7 (relative clause and noun), 8 (proper noun and common noun), 
9 (comparison of superiority), 10 (main verb and auxiliary verb)   
and 19 (purpose clause and main clause), have  strong dependence on  
the parameter 2 (noun and adposition)~\cite{Tsunoda1,Tsunoda2,Ueda}. 
For example our study of the data  gives the  statistical laws 
on cooccurrence with three parameters.  

\begin{enumerate}
\def\labelenumi{(\roman{enumi})} 
\item  If a language has preposition, and if the language is the VO language,  
then the genitive tends to follow the noun. \\ 
If a language has postposition, 
and if it is the OV language, the genitive tends to precede the noun.

\item  If the main verb of a language follows the auxiliary verb, 
and if the relative clause follows the noun, the purpose clause tends 
to follow the noun.\\
If the main verb of a language precedes the auxiliary verb, 
and if the relative clause precedes the noun, the purpose clause tends 
to precede the noun. 
\end{enumerate}

We study  the cooccurrence of 
  word ordering rules of the world's languages by using the idea of the
Ising model. 
It is well known that the spontaneous magnetization 
 can occur in the 
thermodynamic limit only.   
 In a finite lattice, the system  
makes excursions from states with uniformly negative magnetization through 
this intermediate mixed-phase state to states with uniformly positive 
magnetizations~\cite{Binder}.  Our ternary interaction model of a finite 
population  makes  simillar excursions assuming a  mutation to the other type 
for each particle.  
Consider a system of particles of two types $A$ 
 and $B$.  At each step,    three  particles are taken at random.  
If two of the three particles are of the type $A$ ($B$) and one is of the 
type 
$B$ ($A$), the three particles become three 
particles of the type $A$ ($B$)   with probability $p$  
 and the 
three  
particles   of the type $B$ ($A$) with probability  $1-p$. 
We continue  this step sequentially.  
Hence, if $p$ is larger than 1/2, the type of the   majority is  
advantageous 
to  the type of the minority.  If 
  $p$  is less than 1/2,  the type of the minority is 
advantageous.
  The model for the $p$ less than 1/2  might be  a 
simple caricature of the Ising model for the temperature higher  than the 
critical point, although the model does not have 
space variables neglecting the position of each particle.   
 The case of the $p$ higher than 1/2  corresponds to  the Ising model  for
 the 
temperature  lower  than the critical point,  where the majorities 
are advantageous to  the minorities.    
Here we apply the ternary interaction model   to study   
word ordering rules of the world 130 languages~\cite{Tsunoda1}.

A pair of very 
similar word orders may appear just by chance.   
  For instance, the difference  between Tamil and Japanese is as small as the
difference between Italian and Spanish in Tsunoda's word 
order table~\cite{Tsunoda1}.  
Japanese as well as Korean, Tamil and  several other  languages seem to have
a stable structure of word ordering rules.  
We put the value $+1$ for each of the eight parameters `noun and adposition' 
and the above seven parameters for Japanese. 
We put $-1$ for each of the eight parameters for  the languages which have 
the opposite word ordering rules like Thai.  
The changes in the values of the parameters of each language 
may  depend on  the values of
other parameters as can be seen from the above i),  and ii) for three 
parameters. 
It seems that each language in the world fluctuates between these stable two
structures. Hence 
the system of word ordering rules   in  Thai,  or the opposite system like  
Japanese, Korean, Mongolian and others 
seem to be reasonable considering   the above ternary interaction model of  
the majority rule, and 
may be 
explained by  economy of communications  to  avoid misunderstanding.   
Each language in the world has its own history and personality. 
Word ordering rules of each language  may   change at random  like a  
mutation in a population of eight individuals each of which corresponds to 
a parameter of word ordering rules.    
The biological mechanism for performing language may be  ultimately 
functionally 
driven by the need for rapid and efficient communications in real 
time~\cite{Gell}. 
Hawkins~\cite{Hawkins} discusses the relationship between innateness  and 
functional pressures in the explanation of language universals and  argues 
some 
innate processsing mechanism have responded to functional pressure that make 
rapid and efficient communications possible.  This observation may support 
our stochatic model for the change of word ordering rule.
It seems that each language fluctuates between the above stable two 
structures like the Ising model for finite lattice. The    fractal  
structure~\cite{Mandelbrot}  is  useful to understand the  structure of 
sentences~\cite{Tokieda}. 
The parse tree  is convenient to represent the phrase structure rule 
in generative grammar~\cite{Chomsky}. English is a prepositional language and 
Japanese is a postpositonal language. English sentences are usually 
represented by left branching parse trees,  while Japanese sentences are
represented by  left branching parse trees.  
Evolution of word ordering rules or evolution of parse trees, which may be 
closely related with each other,  are  interesting problems and will be 
discussed by using evolutionary game theory~\cite{Nowak}.

\noindent 
{\bf Simulation  and  data}  We consider the numerical values of the above  
eight 
parameters.    
For the parameters of each language that can not
clearly be classified into $+1$ or $-1$, we give  a numerical value
between $+1$ and $-1$ by a digitization of the description in Table 1.   
We take  the summation  of the numerical values of the eight
parameters for each language and divide the value by eight to take the 
arithmetic mean. 
 We take the arithmetic mean for the summation of the existing values of  
parameters when there are missing values.  We call the arithmetic mean as the 
measure of Postposition-ness
Figure 1 shows the histogram of  the measure of ``postposistion-ness''
 considering the eight parameters  for the 130  languages. 

\begin{figure}[hbpt]
\caption{   
The histogram for  the measure of ``postposition-ness'' considering the
eight parameters  on  the 130  languages (from Ueda ans Itoh (1995)).}
\epsfxsize=\textwidth
\epsfbox{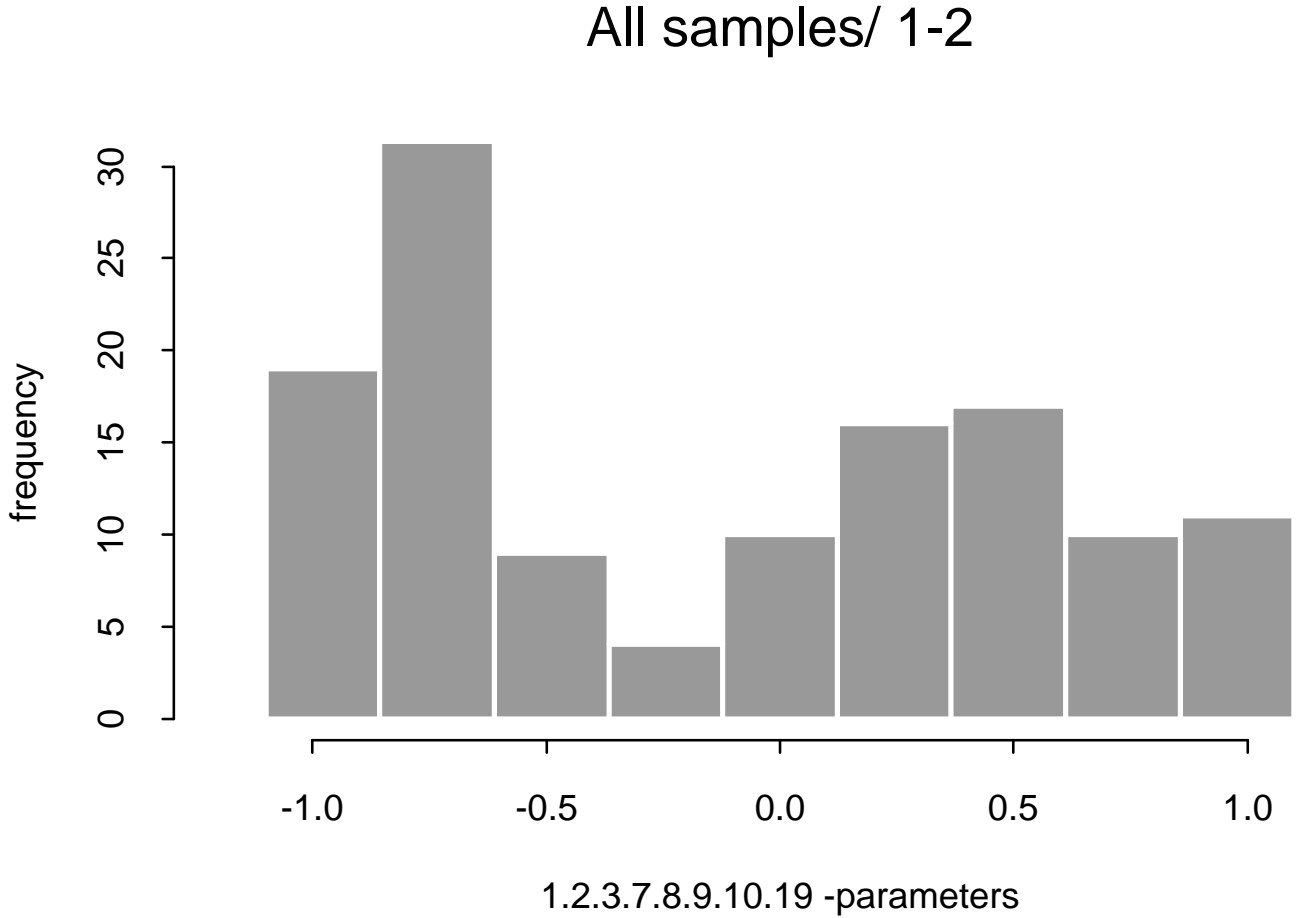}
{\footnotesize \hspace{6em} Figure 1}
\end{figure}

 The changes of word ordering rules could be modeled 
as the above ternary interaction  model of eight particles 1, 2,..., 8,
assuming that $p=1$ and 
the mutations for each particle changing its sign occur at a certain rate 
at the end of each step.  
The trajectory of the number $n_+$  of particles of the type $+1$  is given 
in 
Figure 2a for the first 500 steps, 
assuming a mutation to the opposite sign occurs in the system of eight
particles with probability 0.5 for a particle chosen at random at  each step. 
The histogram  for the numbers of visits has  two modes  as shown 
in Figure 2b.  

\begin{figure}[hbpt]
\caption{
The simulation of the majority rule model. 
a: The trajectory.  b: The histogram with two modes for numbers of visits by the trajectory,   
assuming a mutation to the opposite sign at each step.}
\epsfxsize=\textwidth
\epsfbox{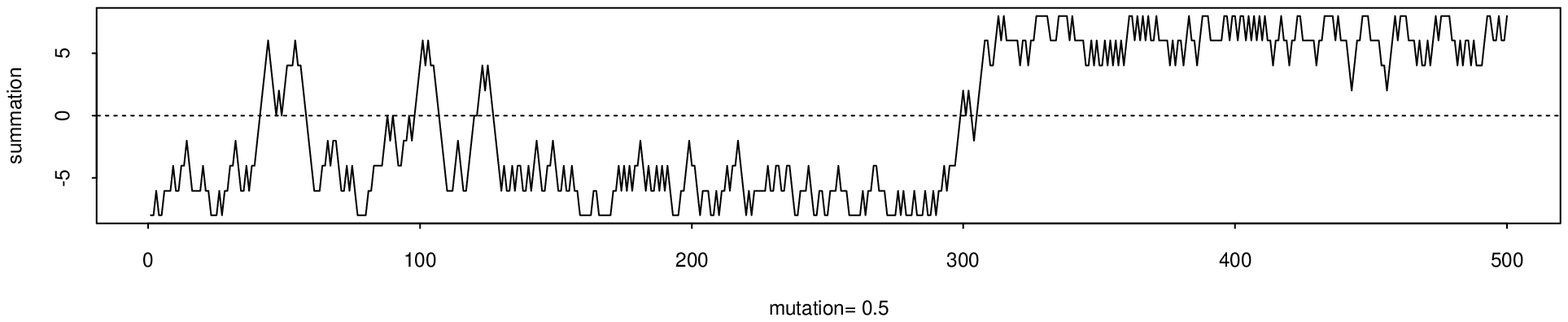}
\epsfxsize=\textwidth
\epsfbox{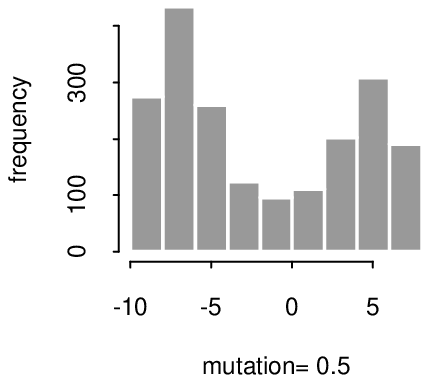}
\end{figure}

We represent  
word ordering rules of each of 130 languages by a vector of 66-dimension. 
We define a Manhattan distance $d$ between two languages as eq. (1) given
later.  
We make  a random walk on the 130 languages to simulate changes of word 
ordering rules.  
We take a language at random at first. 
Then we take the second  language at random from languages within a given 
Manhattan distance $d=0.6$ from the first language. We take the third language
 at random from languages within the distance $d=0.6$ from the second language. 
We continue this random walk and obtain 
the trajectory  given in Figure 3a for the first 500 steps.
The histogram on the number of  visits for  the arithmetic mean also has two
modes 
in  Figure 3b as in Figures 1 and 2b for the first 2000 steps.    
 The states in which 
  the eight parameters have the same sign,  seem  to be relatively 
stable in the random walk.

\begin{figure}[hbpt]
\caption{The random walk on the 130  languages. 
a. The trajectory of the first 500 steps starting from Japanese.   
b. The histogram  of the numbers of visits by the trajectory. } 
\epsfxsize=\textwidth
\epsfbox{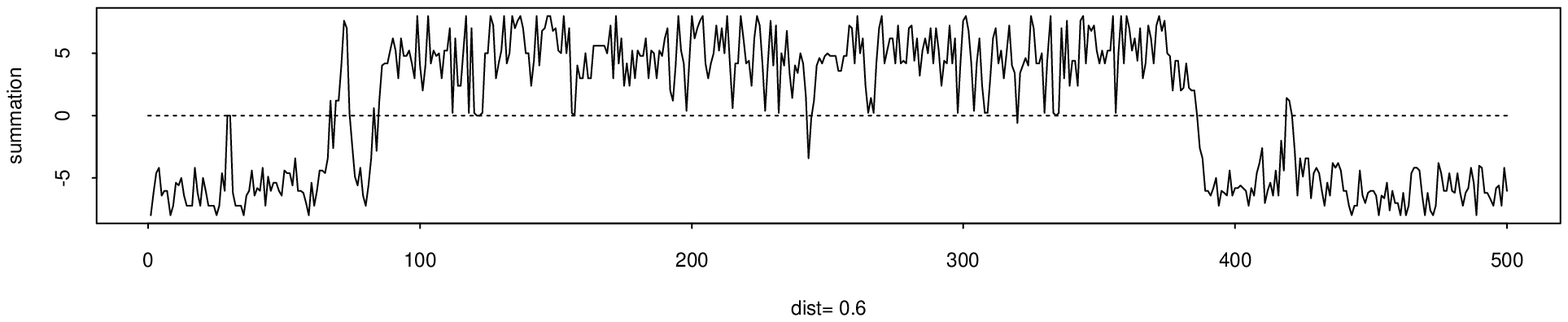}

\vspace{3.5cm}
\epsfxsize=\textwidth
\epsfbox{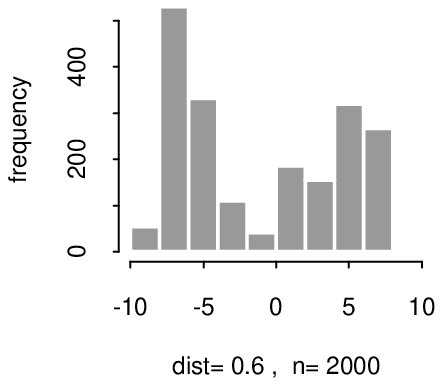}

\end{figure}

\noindent 
{\bf Conclusion}  Our present study on Tsunoda's table may  give  a 
possible   answer to  an aspect of the problem 
how the language universals evolve. 
The word ordering rule may change at random by getting  functional pressure 
that make rapid and efficient communications possible. 
The word ordering rule of each language 
 in the world seems to 
fluctuate between the  two stable  strucures,  a typical postpositional 
language structure  and a typical  prepositional  language structure, 
like the Ising model for finite lattice.

\vspace{2ex}

\noindent
{\bf  Methods for  the digitization of Tsunoda's  Table}

We  make a  66-dimensional vector for each language 
to digitize the word order data. 
In the data by Tsunoda `NA' is used for the case that the author of the
book~\cite{Tsunoda1}  
 had  not enough knowledge to give the description for a
parameter of 
a language at the time of publication of the data.  
Our present study is based on the data by Tsunoda~\cite{Tsunoda1}  
as in the previous study~\cite{Tsunoda2}. Our method can be extended to 
define the distance between two languages including  "NA", applying 
the idea in the S language~\cite{Becker}, although we do not 
explain it in this article.

A numerical value between $0$ and $1$ was assigned to each of the 19 parameters
according to the description of word orders concerned. 
The total of the elements within a parameter should be 1, i.~e.
 $$  \sum_{m=1}^{ n_k} x_{i,k,m} = 1,$$
for  $i=1,\ldots,130$, $k=1,\ldots,19$, 
where $n_k$ is the dimension of parameter $k$ and
$x_{i,k,m}$ is the value of the {\it m}-element of parameter $k$ of 
language $i$.     

We represent the state of the parameter 1 by the 2-dimensional vector for 
the six possible combination of the parameter S,O and V.  
The  OV language contains SOV, OSV and OVS languages.  
The VO language contains SVO, VSO and VOS.    
In the case of Japanese, which has `SOV, etc.', 
$(1, 0)$  is assigned.  In the case of English, which has `SVO',  $(0, 1)$ 
is assigned. 
And also, in the case of Modern Greek, which has `SVO, VSO, etc.',  $(0, 1)$
is assigned. 
For a parameter with a 3-dimensional vector, we represent it by 
 $(1,0,0)$ if the order is the  same to  Japanese. 
If it is the reversed order to Japanese,  we represent it by  $(0,1,0)$. 
Otherwise we represent it by  $(0,0,1)$. 
For example, the parameter 3 (genitive and noun) for English is  `$+,-$'
which is
represented by $(0.6, 0.4, 0)$.
When a parameter of a language has more than one word order possibility,
the ratio is given by the  similarity of the competing possibilities.   
   We use  6-, 5- and 13-dimensional vectors, respectively 
 for the parameters, 11, 13 and 17.  
  For each of the 130 languages, we make a 66-dimensional vector  from the 
   19 vectors  by arranging the coordinates of the 19
vectors from the vector 
for parameter 1 to the vector for parameter 19. 
We give two  examples for the digitization.  
The 19 slashes in the following vector  classify the 66 coordinates into the
19 parameters.

\vspace{1ex}
{\small 
\noindent
Japanese: 
\[
\begin{array}{l}
(1,0 / 1,0,0 / 1,0,0 / 1,0,0 / 1,0,0 / 1,0,0 / 1,0,0 / 1,0 / 1,0,0 / 1,0,0 / \\
 1,0,0,0,0,0 / 1,0,0 / 1,0,0,0,0 / 1,0 /1,0,0 / \\
 1,0 / 1,0,0,0,0,0,0,0,0,0,0,0,0 / 1,0 / 1,0/) 
 \end{array}
 \]
English: 
\[
\begin{array}{l}
 (0,1 / 0,1,0 / 0.6,0.4,0 / 1,0,0 / 1,0,0 / 1,0,0 / 0,1,0 / 0.4,0.6 / 0,1,0
/ 0,1,0 /\\
  0,1,0,0,0,0 / 0.6,0.4,0 / 0,0,1,0,0 / 0,1/ 0,1,0 /\\
  0,1 / 0,0,0,1,0,0,0,0,0,0,0,0,0 / 0.6,0.4 / 0,1/)
\end{array}
\]


We define the distance between two languages $i$ and $j$ i as 
\begin{eqnarray}
d(i,j)  = \frac{1}{19} \sum_{k=1}^{19} \sum_{m \in n_k} \mid x_{i,k,m} -
x_{j,k,m} \mid   \leq 2 .  
\end{eqnarray}

The distances between English and Japanese, between English and Thai, and 
between  
Japanese and Thai are   
1.45,  1.05, and 1.52.
By using the S language~\cite{Becker}  
using the above distances we derived the hierarchical clustering for the
Eurasian
languages (Figure 4). 

\begin{figure}[hbpt]
\caption{The hierarchical clustering for the Eurasian languages using the distance 
metric $d*19$ applying the furthest method in S language (from Tsunoda, Ueda, and Itoh (1995)).}
\epsfxsize=\textwidth
\epsfbox{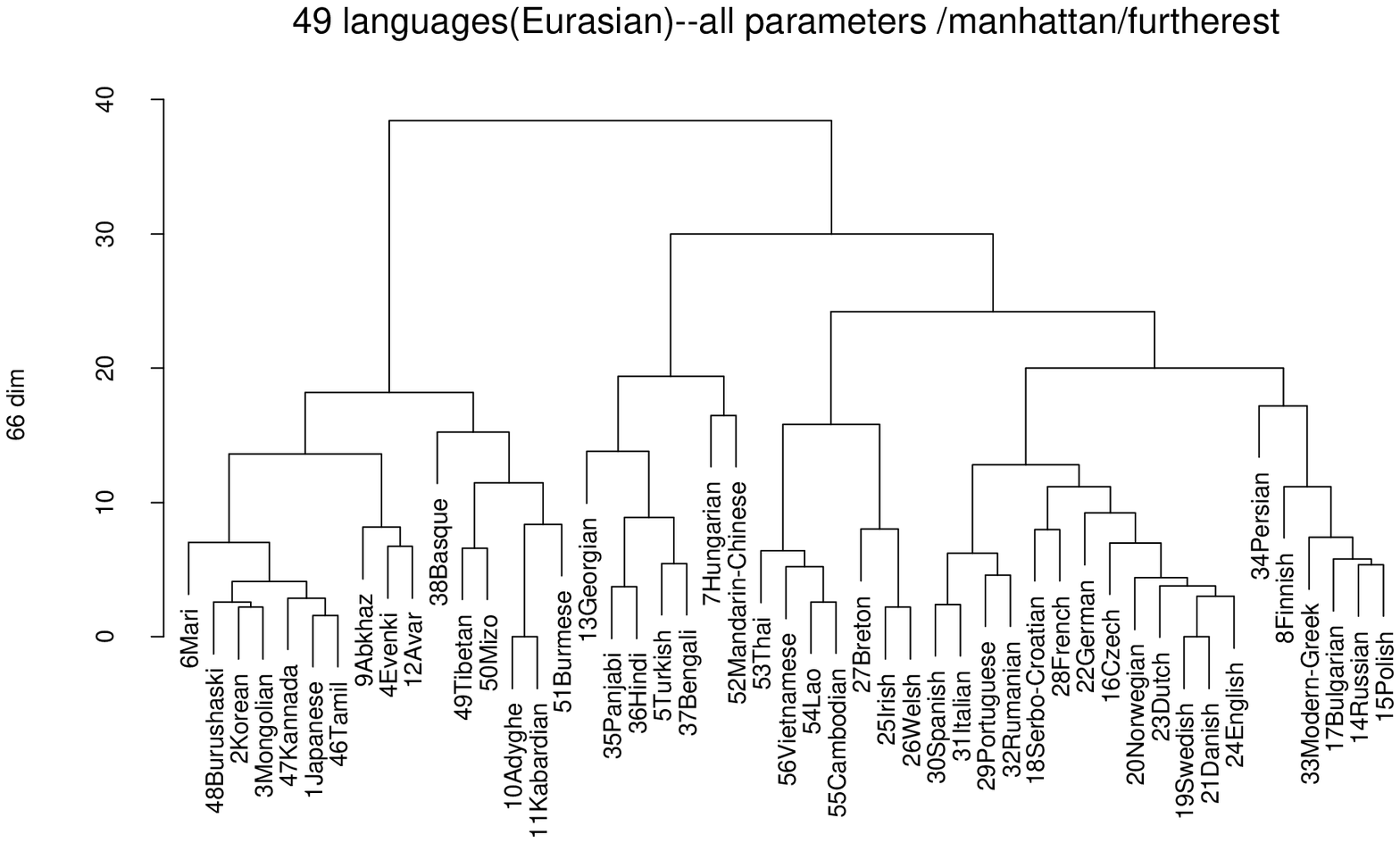}
\end{figure}

The authors are greatful to Professor  Kei Takeuchi for helpful comments. 



\end{document}